\documentclass[english,twoside,a4paper,10pt]{article}

\usepackage[latin1]{inputenc}
\usepackage[T1]{fontenc}
\usepackage{amsmath}
\usepackage{amsfonts}
\usepackage{graphicx}
\usepackage{subfigure}
\usepackage{a4wide}
\usepackage{amssymb}
\usepackage{fancyhdr}
\usepackage{mathrsfs}
\usepackage[toc,page]{appendix}

\begin{document}

\title{\bf Scalar tensor Horndeski models: simple cosmological applications}
\author{ 
Ratbay Myrzakulov\footnote{Email: rmyrzakulov@gmail.com},\,\,\,
Lorenzo Sebastiani\footnote{E-mail address: l.sebastiani@science.unitn.it
}\\
\\
\begin{small}
Department of General \& Theoretical Physics and Eurasian Center for
\end{small}\\
\begin{small} 
Theoretical Physics, Eurasian National University, Astana 010008, Kazakhstan
\end{small}\\
}

\date{}

\maketitle


\begin{abstract}
In this paper, we analyze a simple class of Horndeski scalar-tensor theory. We investigate several cosmological solutions for inflation, bounce scenario and finite future-time singuarities. Perturbations respect to Friedmann-Robertson-Walker metric are studied and discussed in the context of inflation.
\end{abstract}



\tableofcontents
\section{Introduction}

In the last years, modified theories of gravity became very popular due to the possibility to describe a huge variety of cosmological scenarios that standard matter and radiation are not able to reproduce in the classical framework of Einstein's gravity. In modified gravity, the Hilbert-Einstein action of General Relativity, namely the Ricci scalar $R$, is replaced by a more general function of the curvature invariants with the introduction of new freedom degrees in the theory: in the simplest formulation, one deals with a function of the Ricci scalar only (the so called ``$f(R)$-theories'', see Refs.~\cite{R1, R2, R3, R4, R5} for some reviews), but in other models we may work with contractions of the Riemann and/or Ricci tensor, 
interactions with scalar field and so on. At the cosmological level, the aim of this theories is to recover acclerating solutions for the dark energy and inflation (see Ref.~\cite{Staro} or the recent reviews in Refs.~\cite{Odinfrev, myrevinfl}), but also other issues (like the dark matter phenomenology) can be carried out. 

In modified gravity the field equations of the theory are much more complicated respect to the case of General Relativity, leading to fourth order differential equations. However, in 1974,   Horndeski proposed a class of scalar tensor theories where the field equations are at the second order like in General Relativity~\cite{Horn}. The Horndenski Lagrangian is quite involved and contains the interaction of a scalar field with gravity. These theories have been well study and recently many works appeared~\cite{Def, DeFelice, Kob, Kob2, Qiu, Maselli, Zerg, add1, add2, add3, add4}. In this paper, we consider a simple formulation of Horndeski gravity and we investigate several cosmological solutions for inflation, bounce and accelerating universe with the presence of finite future-time singularities.

The paper is organized in the following way. In Section {\bf 2}, the Lagrangian and the Friedmann equations of the model will be presented. In Section {\bf 3}, we will investigate solutions for inflation, bounce scenario and finite future-time singularities. In Section {\bf 4}, we consider perturbations in our solutions for inflation and confront the results with the last Planck data. Conclusions are given in Section {\bf 5}.

We use units of $k_{\mathrm{B}} = c = \hbar = 1$ and 
$8\pi/M_{Pl}^2=1$, where $M_{Pl}$ is the Planck Mass.

\section{Formalism}

The most general class of scalar-tensor theories which bring to second order Equations of Motion (EOMs) reads~\cite{Horn}
\begin{equation}
I=\int_\mathcal M dx^4\sqrt{-g}\left[\frac{R}{2}+\mathcal L_H+L\right]\,,\quad \mathcal{L}_H=\sum_{i=2}^5\mathcal{L}_i\,,
\end{equation}
where $\mathcal{M}$ is the space-time manifold, $g$ is the determinant of the metric tensor $g_{\mu\nu}$, $R$ is the Ricci scalar and represents the Hilbert-Einstein action of General Relativity (GR), $\mathcal L$ is the Lagrangian of the matter contents of the space-time, and $L_H$ collects the higher curvature corrections to GR and is given by
\begin{equation}
\mathcal{L}_2=P(\phi,X)\,,\nonumber
\end{equation}
\begin{equation}
\mathcal{L}_3=-G_3(\phi,X)\Box\phi\,,\nonumber
\end{equation}
\begin{equation}
\mathcal{L}_4=G_4(\phi,X)R+G_{4,X}[(\Box\phi)^2-(\nabla_\mu \nabla_\nu \phi)(\nabla^\mu \nabla^\nu \phi)]\,,\nonumber
\end{equation}
\begin{eqnarray}
\mathcal{L}_5&=&G_5(\phi,X)G_{\mu\nu}(\nabla^\mu \nabla^\nu \phi)-\frac{1}{6}G_{5,X}[(\Box\phi)^3-\nonumber\\&&
3(\Box\phi)(\nabla_\mu \nabla_\nu \phi)(\nabla^\mu \nabla^\nu \phi)+2(\nabla^\mu \nabla_\alpha\phi)(\nabla^\alpha \nabla_\beta \phi)(\nabla^\beta \nabla_\mu \phi)]\,.
\end{eqnarray}
Here, $G_{\mu\nu}:=R_{\mu\nu}-R g_{\mu\nu}/2$ is the usual Einstein's tensor, $R_{\mu\nu}$ being the Ricci tensor, $\phi$ is a scalar field coupled with gravity, and $P(\phi, X)$ and $G_i(\phi, X)$ with $i=3,4,5$ are functions of the scalar field $\phi$ and $X=-g^{\mu\nu}\partial_\mu\phi\partial_\nu\phi/2$.

In Ref.~\cite{Zerg}, the authors investigated a simple class of Horndeski model where a mimetic constraint on the field has been implemented. In our paper, we would like to reconsider such a Lagrangian by using the field as an independent freedom degree of the theory.
The action reads
\begin{equation}
I=\int_\mathcal{M}d^4x\sqrt{-g}\left(\frac{R}{2}+\mathcal{L}-\frac{g^{\mu\nu}\partial_{\mu}\phi\partial_\nu\phi}{2}-V(\phi)\right)+I_H\,,\label{action}
\end{equation}
where
\begin{equation}
I_H=\int_{\mathcal M} d^4 x \sqrt{-g}\left[\alpha\left(
G_{\mu\nu}\nabla^\mu\phi\nabla^\nu\phi\right)
+
\gamma 
\phi G_{\mu \nu} \nabla^\mu \nabla^\nu \phi -\beta \phi  \Box \phi \right]\,,\label{actionH}
\end{equation}
with $\alpha\,,\beta\,,\gamma$ constant coefficients\footnote{If we reintroduce the Planck Mass, the dimension of $\alpha\,,\gamma$ results to be $[\alpha]=[\gamma]=[M_{Pl}^{-4}]$.}.
We note that after integration by part we can also write
\begin{equation}
\int_\mathcal{M}d^4x\sqrt{-g} G_{\mu\nu} \nabla^\mu \nabla^\nu \phi=
\int_\mathcal{M}d^4x\sqrt{-g}\left( -\frac{g^{\mu\nu}\partial_\mu\phi\partial_\nu\phi R}{2} +(\Box \phi)^2-\nabla_\mu \nabla_\nu \phi \nabla^\mu \nabla^\nu \phi\right)\,.
\end{equation}
Then, we see that 
\begin{equation}
\int_\mathcal{M}d^4x\sqrt{-g} G_{\mu\nu} \nabla^\mu\phi \nabla^\nu \phi=
-\int_\mathcal{M}d^4x\sqrt{-g} G_{\mu\nu} \nabla^\mu \nabla^\nu \phi+
\int_\mathcal{M}d^4x\sqrt{-g}G_{\mu\nu}\nabla^{\mu}[\phi\nabla^{\nu}\phi]\,,
\end{equation}
but the second term vanishes after integration,
\begin{equation}
\int_\mathcal{M}d^4x\sqrt{-g}G_{\mu\nu}\nabla^{\mu}[\phi\nabla^{\nu}\phi]=
-\int_\mathcal{M}d^4x\sqrt{-g}\nabla^{\mu}G_{\mu\nu}[\phi\nabla^{\nu}\phi]=0\,.
\end{equation}
We conclude that, in the special case $\alpha=\gamma$, the corresponding contributes disappear from the EOMs.

Let us consider the flat Friedmann-Robertson-Walker (FRW) metric,
\begin{equation}
ds^2=-dt^2+a(t)^2 d{\bf x}^2\,,\label{metric}
\end{equation}
where $a\equiv a(t)$ is the scale factor of the universe and depends on the cosmological time.

In this case, the Friedmann-like equations read (see, for instance, Ref.~\cite{DeFelice}\footnote{For a comparison, put 
$K(\phi, X)=X-V(\phi)$, $G_3=\beta\phi$, $G_4=\alpha X$, $G_5=\gamma\phi$ with $X=\dot\phi^2/2$.}):
\begin{eqnarray}
3H^2(1-3\alpha\dot\phi^2+3\gamma\dot\phi^2)=\rho+\frac{\dot\phi^2}{2}+V(\phi)-\beta\dot\phi^2\,,\label{EOM1}
\end{eqnarray}
\begin{eqnarray}
-(2\dot H+3H^2)&=&p+\frac{\dot\phi^2}{2}-V(\phi)-\beta\dot\phi
+\alpha\dot\phi^2(3H^2+2\dot H)
-6\alpha H^2\dot\phi^2-4\alpha H\dot\phi\ddot\phi-4\alpha\dot H\dot\phi^2\nonumber\\&&
+\gamma(2\dot H\dot\phi^2+4H\dot\phi\ddot\phi+3H^2\dot\phi^2)\,,\label{EOM2}
\end{eqnarray}
where the dot denotes the derivative with respect to the time and $\rho, p$ are the energy density and the pressure of the universe contents, respectively, which satisfy the relation
\begin{equation}
\dot\rho+3H(\rho+p)=0\,.\label{conslawmatter}
\end{equation}
Finally, the variation of the action respect to $\phi$ leads to the continuity equation for the field
\begin{equation}
\ddot\phi(1-2\beta+6\alpha H^2-6\gamma H^2)
+3H\dot\phi(1-2\beta+6\alpha H^2-6\gamma H^2)
+12H\dot H\dot\phi(\alpha -\gamma)=-V_\phi(\phi)\,,\label{conslaw}
\end{equation}
where the pedex ``$\phi$'' denotes the derivative respect to $\phi$.
As we observed above, in the special case $\alpha=\beta$ the theory is recasted in the background of General Relativity. 

\section{Cosmological solutions}

We introduce the $e$-folds number $N$ as
\begin{equation}
N=\log\left[\frac{a(t_0)}{a(t)}\right]\,,\label{N}
\end{equation}
where $a(t_0)$ is the scale factor at the fixed time $t_0$. Thus, since $dN=-Hdt$, Equations~(\ref{EOM1}, \ref{conslawmatter}, \ref{conslaw}) lead to
\begin{equation}
3H^2 +9\tilde\alpha H^4\phi'^2=\rho+\tilde\beta H^2\frac{\phi'^2}{2}+V(\phi)\,,\label{EOM1bis}
\end{equation}
\begin{equation}
\rho'-3(\rho+p)=0\,,\label{consmatter}
\end{equation}
\begin{equation}
\phi''(\tilde\beta-6\tilde\alpha H^2)+\frac{H'\phi'}{H}(\tilde\beta-18\tilde\alpha H^2)
-3\phi'(\tilde\beta-6\tilde\alpha H^2)=-\frac{V_\phi(\phi)}{H^2}\,.\label{conslawbis}
\end{equation}
In the above expressions, the prime denotes the derivative with respect to $N$ and
\begin{equation}
\tilde\alpha=\gamma-\alpha\,,\quad\tilde\beta=1-2\beta\,.
\end{equation}
Let us see some cosmological application of the model.

\subsection{Solutions for inflation\label{inflsol}}

In 1981 Guth~\cite{Guth} and Sato~\cite{Sato} suggested that the universe underwent a period of strong accelerated expansion after the Big Bang, namely the inflation. Thanks to inflation, one may explain the thermalization of observable universe and can solve several problems related with the initial conditions of Friedmann universe~\cite{Linde, revinflazione}.
The scalar tensor theories of gravity are quite interesting in the context of inflation, and lead to the possibility to reproduce the early-time acceleration by using a scalar field subjected to a potential~\cite{chaotic}. Here, we will see how we can obtain this scenario by considering the additional higher-order corrections in our Lagrangian. Later, we will discuss the results by investigating the spectral index and the tensor-to-scalar ratio of the primordial flactuations.

In the analysis of inflation, the indtroduction of the $e$-folds in (\ref{N}) results to be particularly useful if we set $t_0$ as the time at the end of the eraly-time acceleration, when the universe enters in the radiation-dominated era with the rehating processes for particle production. Thus, the total $e$-folds is defined by $N$ at the beginning of inflation, when $t=t_\text{i}$,
\begin{equation}
\mathcal N=N|_{t=t_\text{i}}=\log\left[\frac{a(t_0)}{a(t_\text{i})}\right]\,.\label{NN}
\end{equation}
In order to get the thermalization of observable universe, we need $55<\mathcal N<65$. During inflation the Hubble parameter is almost a constant and near to the Planck scale. The slow-roll approximation takes into account the fact that in this phase the field slowly changes such that $|H^2 \phi'^2|\ll V(\phi)$ and $|\phi''|\ll H |\phi'|$. We also may assume
\begin{equation}
 H^2 \phi'^2\ll\frac{1}{|\tilde\alpha|}\,,\label{slowalpha}
\end{equation}
where we remember that the natural scale of $\tilde\alpha$ is $|M_{Pl}^{-4}|$. During inflation, 
the slow-roll parameter
\begin{equation}
\epsilon=\frac{H'}{H}\,,\label{epsilon}
\end{equation}
has to be positive and very small. Inflation ends when $\epsilon=1$.
Thus, if we neglect the contribute of matter, Eqs.(\ref{EOM1bis}, \ref{conslawbis}) read
\begin{equation}
3H^2\simeq V(\phi)\,,\quad
\phi'\simeq\frac{V_\phi(\phi)}{3H^2(\tilde\beta-6\tilde\alpha H^2)}\,.\label{eqslowroll}
\end{equation}
Typically, the field at the beginning of inflation is negative and its magnitude is very large, while at the end of inflation the field tends to vanish with the potential.
Therefore, since we need $\phi'<0$ and $V'(\phi)<0$, we must require 
\begin{equation}
0<\tilde\beta-6\tilde\alpha H_\text{dS}^2\,,
\end{equation}
as a general feature of the model.

Let us have a look for some simple solutions for inflation. To start, we may choose
\begin{equation}
H^2=H_0^2(N+1)\,,\label{H1}
\end{equation}
where $H_0$ is the Hubble parameter at the end of inflation. Thus, from (\ref{eqslowroll}) one has
\begin{equation}
V(\phi)\simeq 3H_0^2(N+1)\,,\quad \phi'\simeq \frac{V_\phi(\phi)}{3H^2(\tilde\beta-6\alpha H^2)}\,.\label{esempio1}
\end{equation}
The de Sitter solution for inflation reads
\begin{equation}
H_\text{dS}^2=H_0^2(\mathcal N+1)\,,
\end{equation}
with $\mathcal N$ the total $e$-folds left from inflation as in (\ref{NN}). We immediatly note that 
\begin{equation}
\epsilon\simeq\frac{H_0^2}{2H_\text{dS}^2}\simeq \frac{1}{2(\mathcal N+1)}\ll 1\,,
\end{equation}
and the slow-roll approximation holds true.
The second equation in~(\ref{esempio1}) can be solved respect to $\phi$ in two limiting cases. At first, we may take $\tilde\alpha=0$, such that 
\begin{equation}
\phi\simeq\phi_\text{0}-2\sqrt{1+N}\sqrt{\frac{1}{\tilde\beta}}
\,,
\label{25}
\end{equation}
and one can reconstruct the potential which leads to this kind of inflation as,
\begin{equation}
V(\phi)=\frac{3H_0^2}{4}\tilde\beta(\phi-\phi_0)^2\,.
\label{ex1}
\end{equation}
Here we remember that if we want to reintroduce the Planck Mass, we must shift $H_0^2\rightarrow 8\pi H_0^2/M_{Pl}^2$.

Now we could take the limit $\tilde\beta=0$. Thus, from (\ref{esempio1}),
\begin{equation}
\phi\simeq\phi_\text{0}-\frac{1}{H_0\sqrt{6}}\sqrt{\frac{1}{-\tilde\alpha}}\log[N+1]
\,,
\end{equation}
with $\tilde\alpha<0$,
and the potential reads
\begin{equation}
V(\phi)\simeq 3H_0\text{e}^{(\phi_0-\phi)\sqrt{-6H_0^2\tilde\alpha}}\,.
\label{ex2}
\end{equation}
A generalization of (\ref{H1}) is given by
\begin{equation}
H^2=H_0^2(\mathcal N+1)^\lambda\,,\quad 0<\lambda\,,
\end{equation}
where $\lambda$ is a positive paramter. Then,
\begin{equation}
V(\phi)\simeq 3H_0^2(N+1)^\lambda\,,\quad \phi'\simeq \frac{V_\phi(\phi)}{3H^2(\tilde\beta-6\alpha H^2)}\,,\label{esempio1}
\end{equation}
with $H_\text{dS}^2=H_0^2(\mathcal N+1)^\lambda$ and $\epsilon\simeq\lambda/(2(\mathcal N+1))\ll 1$. By taking the limit $\tilde\alpha=0$, one has
\begin{equation}
\phi\simeq\phi_0-2\sqrt{1+N}\sqrt{\frac{\lambda}{\tilde\beta}}\,,\quad
V(\phi)=\frac{3H_0^2}{(4\lambda)^\lambda}\tilde\beta^{\lambda}(\phi_0-\phi)^{2\lambda}\,.
\label{ex1bis}
\end{equation}
On the otehr hand, in the limit $\tilde\beta=0$ one obtains
\begin{equation}
\phi\simeq\phi_\text{0}\pm\frac{\sqrt{2}}{H_0(\lambda-1)\sqrt{3}}\sqrt{\frac{\lambda}{-\tilde\alpha}}(1+N)^{\frac{1-\lambda}{2}}\,,\quad
V(\phi)=
3H_0^2\left[-\frac{3\tilde\alpha H_0^2}{2\lambda} (\lambda-1)^2 (\phi-\phi_0)^2\right]^{\frac{\lambda}{1-\lambda}}\,,
\label{ex1tris}
\end{equation}
where the sign minus corresponds to $1<\lambda$ and the sign plus to $0<\lambda<1$.

An other example is given by the following behaviour of the Hubble parameter,
\begin{equation}
H^2=H_\text{dS}^2\left[1-\frac{1}{(N+1)}\right]^2\,,
\end{equation}
where $H_\text{dS}$ is the Hubble parameter during inflation when $1\ll N$. Therefore,
\begin{equation}
V(\phi)\simeq 3H_\text{dS}^2\left[1-\frac{1}{(N+1)}\right]^2\,,\quad \phi'\simeq \frac{V_\phi(\phi)}{3H_\text{dS}^2(\tilde\beta-6\tilde\alpha H_{\text{dS}}^2)}\,.\label{esempio3}
\end{equation}
Now the $\epsilon$ slow-roll parameter (\ref{epsilon}) reads
\begin{equation}
\epsilon\simeq\frac{1}{(\mathcal N+1)^2}\ll 1\,,
\end{equation}
and it is quadratic respect to the inverse of the $e$-folds (this fact will imply a very small tensor-to-scalar ratio at the end of inflation). The field and the potential are derived as
\begin{equation}
\phi\simeq\phi_0-\frac{\sqrt{2}}{\sqrt{\tilde\beta-6\tilde\alpha H_\text{dS}^2}}\log (N+1)\,,\quad
V(\phi)\simeq 3H_\text{dS}^2\left[1-\text{e}^{(\phi-\phi_0)\sqrt{(\tilde\beta-6\tilde\alpha H_\text{dS}^2)/2}}\right]\,,
\label{ex3}
\end{equation}
$\phi_0$ being the value of the field at the end of inflation. 

As a last example, we will consider 
\begin{equation}
H^2=H_\text{dS}^2\left[1-\frac{1}{(N+1)^{\zeta}}\right]\,,\quad 1<\zeta\,,
\end{equation}
and
\begin{equation}
V(\phi)\simeq 3H_\text{dS}^2\left[1-\frac{1}{(N+1)^\zeta}\right]\,,\quad \phi'\simeq \frac{V_\phi(\phi)}{3H_\text{dS}^2(\tilde\beta-6\tilde\alpha H_{\text{dS}}^2)}\,.\label{esempio3}
\end{equation}
The $\epsilon$ slow-roll parameter (\ref{epsilon}) reads
\begin{equation}
\epsilon\simeq\frac{\zeta}{2(\mathcal N+1)^{\zeta+1}}\ll 1\,,
\end{equation}
which justifies our choice $1<\zeta$, since for $0<\zeta<1$ this parameter will make too large the tensor-to-scalar ratio of the model (see \S~\ref{pert}).
Finally one gets
\begin{equation}
\phi\simeq\phi_\text{i}+\frac{1}{\Phi}\frac{1}{(N+1)^{(\zeta-1)/2}}\,,\quad
V(\phi)\simeq 3H_\text{dS}^2\left[1-(-\Phi(\phi-\phi_\text{i}))^\frac{2\zeta}{\zeta-1}\right]\,,\label{ex5}
\end{equation}
where $\phi_\text{i}$ is the value of the field at the beginning of inflation and
\begin{equation}
\Phi=\sqrt{\frac{\tilde\beta-6\tilde\alpha H_\text{dS}^2}{\zeta}}\left(\frac{\zeta-1}{2}\right)\,.
\end{equation}

\subsection{Bounce and finite future-time singularity solutions}

The bounce scenario, where a cosmological contraction is followed by an
expansion at a finite time, is interesting to analyze as an alternative description respect to the Big Bang theory:
instead from an initial singularity, the universe emerges from a cosmological bounce with an accelerating expansion (see Ref.~\cite{Novello} for a review). In terms of the cosmological time the Hubble parameter is described by
\begin{equation}
H=h_0(t-t_0)^{2n+1}\,,\quad 0<h_0\,,\label{Hbounce}
\end{equation} 
where $h_0$ is a positive dimensional constant, $n$ a natural number and $t_0$ the (positive) time of the bounce: for $t<t_0$ the Hubble parameter is negative and describes a contracting universe, while for $t_0<t$ the Hubble parameter is positive and describes an expanding universe. 

The study of the bounce is similar to the one of the finite future-time singularities. 
A future-time singularity appears if exists some finite time for which the Hubble parameter, the effective energy density or pressure of the universe or -more simplicity- some derivatives of the Hubble parameter diverge~\cite{classificationSingularities}.
This kind of solutions brings the universe to accelerate and is often studied in the context of the dark energy issue.
The Hubble parameter assumes the form
\begin{equation}
H=\frac{h_0}{(t_0-t)^\lambda}\,,\quad 0<h_0\,,\label{Hsing}
\end{equation} 
where $h_0$ is a positive dimensional parameter again and $\lambda$ is a real number. If $0<\lambda$
the Hubble parameter diverges at the time $t=t_0$, otherwise some of its deriveatives may become singular. 

For $\lambda=1$ we obtain the Big Rip~\cite{Caldwell}, realized in General Relativity ($\tilde\alpha=\tilde\beta=0$) by phantom fluids with $p=\omega\rho$, $\omega<-1$ in (\ref{consmatter}), such that in this paper we will omit to speak about this particular case. \\ 
\\
For solution (\ref{Hbounce}) we find
\begin{equation}
a(t)=a(t_0)\exp\left[\frac{h_0(t-t_0)^{2+2n}}{(2+2n)}\right]\,,\quad N=
-\frac{h_0(t-t_0)^{2+2n}}{2+2n}\,,
\end{equation}
where $N$ is the $e$-folds left to the bounce at $a(t)=a(t_0)$. Note that $N<0$, since $a(t_0)$ is the minimal radius reached by the universe at the end of the contraction phase. Thus, we have
\begin{equation}
|H|=\tilde h(-N)^q\,,\quad \tilde h_0=h_0^{\frac{1}{2+2n}}(2+2n)^{\frac{1+2n}{2+2n}}\,,
\quad q=\frac{1+2n}{2+2n}\,.
\end{equation}
If we ignore the matter contributes, some simple solution can be inferred from (\ref{EOM1bis}) and (\ref{conslawbis}) when $|N|\ll 1$, namely near to the bounce. For example, for $\phi=\phi_1 N$, 
with $\phi_1=(\tilde\beta\pm\sqrt{24\tilde\beta+\tilde\beta^2})/(2\tilde\beta)$,
one has that the potential
\begin{equation}
V(N)=\frac{\tilde h_0^2}{2}(-N)^{2q}(6-\tilde\beta\phi_1^2)\rightarrow
V(\phi)=\frac{\tilde h_0^2}{2}\left(-\frac{\phi}{\phi_1}\right)^{2q}(6-\tilde\beta\phi_1^2)\,,
\end{equation} 
is a solution of (\ref{EOM1bis}) and (\ref{conslawbis}) when $N\rightarrow 0$ and $\phi\rightarrow 0$ (the first expression has been derived for $0<H$, the second one is the generalization of the solution to $H\lessgtr 0$, since $\phi$ may be either positive or negative).\\
\\
For solution (\ref{Hsing}), when $\lambda\neq 1$, we have
\begin{equation}
a(t)=a(t_0)\exp\left[\frac{h_0(t_0-t)^{1-\lambda}}{(\lambda-1)}\right]\,,\quad N=
-\frac{h_0(t_0-t)^{1-\lambda}}{\lambda-1}\,,
\end{equation}
where $N$ is the $e$-folds left to $a(t_0)$.  If $\lambda< 1$, $a(t_0)$ corresponds to the scale factor when singularity occurs and $N$ is always positive and goes to zero at $t=t_0$. If $1<\lambda$, the scale factor diverges at the time of singularity: in this case, $a(t_0)$ corresponds to $1\ll (t_0-t)$ and the $e$-folds is always negative and diverges as $N\rightarrow-\infty$ at $t=t_0$.
We get
\begin{equation}
H=\frac{\tilde h_0}{(-N(\lambda-1))^q}\,,\quad \tilde h_0=h_0^{\frac{2-\lambda}{1-\lambda}}\,,\quad q=\frac{\lambda}{1-\lambda}\,.
\end{equation}
If $1<\lambda$ such that $q<0$, a simple solution near to the singularity ($N\rightarrow-\infty$) can be found for $\phi=\phi_1(-N(\lambda-1))^2$, $\phi_1=1/(2(2q-1)(\lambda-1)^2)$ and is realized by the potential
\begin{equation}
V(N)=36\tilde h^4\tilde\alpha(\lambda-1)^4 N^2(-N(\lambda-1))^{-4q}\phi_1^2\rightarrow
V(\phi)=36\tilde h^4\tilde\alpha(\lambda-1)^2 (\phi/\phi_1)^{1-2q}\phi_1^2\,.
\end{equation}
In this case, the Hubble parameter, the scale factor, and the effective energy density and pressure of the universe diverge at the time $t=t_0$.
On the other hand, if $\lambda<0$ and $q<0$, a (soft) singularity in the derivatives of the Hubble parameter appears when $N=0$. In this case, near to the singularity, for $\phi=-6 N/\tilde\beta$ the potential
\begin{equation}
V(N)=\frac{3\tilde h^2(\tilde\beta-6)}{\tilde\beta}(N(1-\lambda))^{-2q}\rightarrow
V(\phi)=\frac{3\tilde h^2(\tilde\beta-6)}{\tilde\beta}(\tilde\beta\phi(\lambda-1)/6)^{-2q}\,,
\end{equation}
realizes the solution (similarly to the bounce case). The strongest singularities of this type occur when $-1<\lambda<0$ such that the first derivative of the Hubble parameter and therefore the effective pressure of the universe diverge.
Finally, if $0<\lambda<1$ and $0<q$, the singularity still occurs at $N=0$. When $1\ll q$, for $\phi=N/2$ we reconstruct the potential
\begin{equation}
V(N)=\frac{9\tilde h^4\tilde\alpha}{4}(N(1-\lambda))^{-4q}\rightarrow
V(\phi)=\frac{9\tilde h^4\tilde\alpha}{4}(2\phi(1-\lambda))^{-4q}\,.
\end{equation}
In this case the Hubble parameter but not the scale factor diverges at the time of singularity.

We conclude the chapter with some remarks. Usually, bounce cosmology prohibits future singularity in the same model, or it stops to be bounce. The example where it is still possible (but only with the softest singularities, namely $\lambda<-1$ in (\ref{Hsing})) is given in Ref.~\cite{Odbs}. An other interesting scenario may be the singular inflation (see the recent works in Refs.~\cite{is1, is2}).

\section{Cosmological perturbations\label{pert}}

Let us discuss the perturbations around the FRW space-time (\ref{metric}) in our model during inflation following Refs.~\cite{Def, DeFelice}. 
By taking into account the scalar perturbations $\alpha(t, {\bf x})\,,\psi(t, {\bf x})$ and $\zeta\equiv\zeta(t,{\bf x})$ inside the FRW metric we get
\begin{equation}
ds^2=-[(1+\alpha(t, {\bf x}))^2-a(t)^{-2}\text{e}^{-2\zeta(t, {\bf x})}(\partial \psi(t,{\bf x}))^2]dt^2+2\partial_i\psi
(t,{\bf x})dt dx^i+a(t)^2
\text{e}^{2\zeta(t, {\bf x})}d{\bf x}\,.
\end{equation}
By plugging this metric into the action (\ref{action})--(\ref{actionH}) and by using the EOMs, we derive
\begin{equation}
I=\int_\mathcal{M}dx^4 a^3\left[A\dot\zeta^2-\frac{B}{a^2}(\nabla\zeta)^2\right]\,,\label{pertaction}
\end{equation}
where 
\begin{eqnarray}
A &\equiv&\frac{\dot\phi^2(1+\tilde\alpha\dot\phi^2)(-6H^2\tilde\alpha+\tilde\beta+\tilde\alpha(18H^2\tilde\alpha+\tilde\beta)\dot\phi^2)}{2(H+3H\tilde\alpha\dot\phi^2)^2}\,,
\nonumber\\
B &\equiv&\frac{1}{(H+3H\tilde\alpha\dot\phi^2)^2}\times\nonumber\\
\hspace{-1.5cm}&&\left(
-(1+3\tilde\alpha\dot\phi^2)(-4H^2\tilde\alpha^2\dot\phi^4+\dot H(1+\tilde\alpha\dot\phi^2)^2)
+2H\tilde\alpha\dot\phi(1+\tilde\alpha\dot\phi^2)(-1+3\tilde\alpha\dot\phi^2)\ddot\phi
\right)\,.\nonumber\\&&
\end{eqnarray}
Thus, one identifies the squared sound speed as
\begin{equation}
c_s^2\equiv\frac{B}{A}=\frac{
\left(-2(1+3\tilde\alpha\dot\phi^2)(-4H^2\tilde\alpha^2\dot\phi^4+\dot H(1+\tilde\alpha\dot\phi^2)^2)
+4H\tilde\alpha\dot\phi(1+\tilde\alpha\dot\phi^2)(-1+3\tilde\alpha\dot\phi^2)\ddot\phi\right)}
{\left(\dot\phi^2(1+\tilde\alpha\dot\phi^2)\left(-6H^2\tilde\alpha+\tilde\beta+
\tilde\alpha(18H^2\tilde\alpha+\tilde\beta)\dot\phi^2\right)\right)}\,.
\end{equation}
In order to avoid ghost and instabilities under the scalar perturbations we must require $0<A, B$.
In terms of the $e$-folds left to the end of inflation (\ref{N}) and
by taking into account (\ref{slowalpha}), namely $|\dot\phi^2|\ll 1/|\tilde\alpha|$, we obtain
\begin{equation}
c_s^2=\frac{2\left(H'+2H^3\tilde\alpha\phi'(2H^2\tilde\alpha\phi'^3+\phi'')\right)}{(H\tilde\beta-6H^3\tilde\alpha)\phi'^2}\,.
\end{equation}
When $\tilde\alpha=0$, $\tilde\beta=1$ (chaotic scalar field inflation) we recover $c_s=-2\dot H/\dot\phi^2$ and by using (\ref{eqslowroll}) one has $c_s^2=1$, but in general $c_s\neq 1$ is allowed.
In the cases of the models described by the potentials (\ref{ex1}, \ref{ex2}, \ref{ex1bis}, \ref{ex1tris}, \ref{ex3}, \ref{ex5}), we still find (in the limit $N\rightarrow\mathcal N$, $1\ll\mathcal N$)
$c_s^2=1$.

Let us return to the effective action (\ref{pertaction}).
If we decompose $\zeta(\eta, {\bf x})$ in the Fourier modes $u(\eta, {\bf k})$, $ {\bf k}$ being the wave number of the mode (the wavelenght is given by $\lambda\sim a/k$), 
\begin{equation}
\zeta(\eta, {\bf x})=\frac{1}{(2\pi)^3}\int d^3k \text{e}^{i {\bf k}{\bf x}}[u(\eta, {\bf k}) a({\bf k})+u^*(\eta, -{\bf k})a^+(-{\bf k})]\,,\label{start}
\end{equation}
with $a({\bf k})^+\,, a({\bf k})$ the creation/annihilation operators associated to ${\bf k}$, we obtain the quantum action for perturbations,
\begin{equation}
I=\int d\eta dx^3\left(\left(\frac{d v}{d\eta}\right)^2-c_s^2(\nabla v)^2+\frac{d^2 z}{d\eta^2}\frac{u^2}{z}\right)\,.\label{action2}
\end{equation}
In these expressions, $d\eta=dt/a$ is the conformal time, we have introduced the sound speed $cs\equiv B/A$, and 
\begin{equation}
v\equiv v(\eta, {\bf x})=z(\eta) u(\eta, {\bf x})\,,\quad z\equiv z(\eta)=a\sqrt{2A}\,.
\end{equation}
From (\ref{action2}) we get
\begin{equation}
\frac{d^2 v}{d\eta^2}-c_s^2\bigtriangleup v-\frac{1}{z}\frac{d^2 z}{d\eta^2}=0\,.
\end{equation}
Note that this equation corresponds to
\begin{equation}
\frac{d}{dt}(a^3 A\dot\zeta)-(B a)\bigtriangleup\zeta=0\,,
\end{equation} 
namely the variation of (\ref{pertaction}) respect to $\zeta$. 
By assuming that the spatial dependence of $u(\eta, {\bf x})$, and therefore of $v(\eta, {\bf x})$, is given by $\exp[i {\bf k}{\bf x}]$, one has
\begin{equation}
\frac{d^2 v}{d\eta^2}+\left(k^2c_s^2-\frac{1}{z}\frac{d^2 z}{d\eta^2}\right)v=0\,.
\end{equation}
The behaviour of $(d^2 z/dz^2)/z$ when $\tau\rightarrow-\infty$ is given by $(d^2 z/dz^2)/z\simeq 2/\tau^2$ and therefore, if we take $a\simeq -1/(H\tau)$ for (quasi) de Sitter expansion, the solution of this equation in the asymptotic limit reads~\cite{DeFelice}
\begin{equation}
v(\eta, {\bf k})\simeq c_0\frac{a H}{(c_s k)^{3/2}}\text{e}^{-i c_s k\eta}(1+i c_s k\eta)\,.
\end{equation}
Here, the constant $c_0$ must be fixed in order to recover the  Bunch-Davies vacuum state $v(\eta, {\bf k})=\exp[-i c_s k\eta]/\sqrt{2c_s\kappa}$ in the asymptotic past $k\eta\rightarrow-\infty$, such that 
$v(\eta, {\bf k})$ obeys to the commutation relation 
\begin{equation}
\left[ v(\eta, {\bf k}) \frac{d v(\eta, {\bf k})^*}{d\eta}, \frac{ d v(\eta, {\bf k})}{d\eta}v(\eta, {\bf k})^*\right]=i\,,
\end{equation}
as a consequence of the bosonic commutation relations on $a({\bf k})^+$ and $a({\bf k})$ in (\ref{start}).
Thus, it must be $c_0=i/\sqrt{2}$ and finally
\begin{equation}
u(\eta, {\bf k})\equiv \frac{v(\eta, {\bf k})}{a\sqrt{2A}}=i\frac{H}{2\sqrt{A}(c_s k)^{3/2}}\text{e}^{-i c_s k\eta}(1+i c_s k\eta)\,.
\end{equation}
The dimensionless variable which represents the amplitude of the fluctuations on the scale $\lambda\sim1/\kappa$ is derived by the power spectrum
\begin{equation}
\mathcal P_{\mathcal R}\equiv\frac{|\zeta_k|^2 k^3}{2\pi^2}=\frac{H^2}{8\pi^2 c_s^3 A}\,,
\end{equation}
and should be evaluated at the crossing of the Hubble horizon when $k= a H/c_s$.
In this expression, $|\zeta_\kappa|=|u(\eta, k)|$ corresponds to the vacuum expectation values of the modes
of $<0|\zeta(\eta, {\bf x})\zeta(\eta, {\bf y})|0>$ inside (\ref{start}). The spectral index is inferred as
\begin{equation}
1-n_s=-\frac{d\ln \mathcal P_{\mathcal R}}{d \ln k}|_{k=a H/c_s}=2\epsilon+\eta_{sF}+s\,,
\end{equation}
where we make use of the notation of Ref.~\cite{DeFelice},
\begin{equation}
\epsilon=-\frac{\dot H}{H^2}\,,\quad \eta_{sF}=\frac{\dot \epsilon_s F+\epsilon_s\dot F}{H (\epsilon_s F)}\,,\quad s=\frac{\dot c_s}{H c_s}\,,\quad \epsilon_s=\frac{A c_s^2}{F}\,,
\end{equation}
with
\begin{equation}
F=1+\alpha\dot\phi^2\,.
\end{equation}
In a similar way, by taking into account the tensor perturbations in the FRW space-time, one obtains the related tensor-to-scalar ratio
\begin{equation}
r=16 c_s\epsilon_s\,.
\end{equation}
In terms of the $e$-folds (\ref{N}) with (\ref{slowalpha}), we obtain for our Lagrangian
\begin{equation}
1-n_s=-\frac{2(2(\tilde\beta-9\tilde\alpha H^2)H'^2\phi'+H(\tilde\beta-6\tilde\alpha H^2)H'\ddot\phi+H(-\tilde\beta+6\tilde\alpha H^2)\phi'(H''+2\tilde\alpha H^3\phi'\phi'''))}{
H(-\tilde\beta+6\tilde\alpha H^2)\phi'(H'+2\tilde\alpha H^3\phi'\phi'')}\,,
\end{equation}
\begin{equation}
r=16\sqrt{2}(\tilde\beta-6\tilde\alpha H^2)\phi'^2\left(
\frac{H'+2\tilde\alpha H^3\phi'\phi''}{H(\tilde\beta-6\tilde\alpha H^2)\phi'^2}\,,
\right)^{3/2}\,,
\end{equation}
that must been evaluated at $N\simeq \mathcal N$.
We observe that for $\tilde\beta=1$ and $\tilde\alpha=0$, by using (\ref{eqslowroll}) one finds
the usual relations for chaotic inflation
\begin{equation}
1-n_s=3\left(\frac{V_\phi(\phi)}{V(\phi)}\right)^2-2\left(\frac{V_{\phi\phi}(\phi)}{V(\phi)}\right)\,,\quad
r=8\left(\frac{V_\phi(\phi)}{V(\phi)}\right)^2\,.
\end{equation}
Let us calculate the spectral index and the tensor-to-scalar ratio of the inflationary solutions presented in \S\ref{inflsol}.
For the potentials in (\ref{ex1}) and (\ref{ex2}) one has
\begin{equation}
1-n_s=\frac{2}{\mathcal N}\,,\quad r=\frac{8}{\mathcal N}\,.\label{cucu}
\end{equation}
These indexes are the same of chaotic inflation with potential $V(\phi)\propto (-\phi)^2$. Thus, the spectrum of perturbations is independent on the choices of $\tilde\beta$ and $\tilde\alpha$.
The last cosmological data~\cite{Planck} constraint these quantities as
$n_{\mathrm{s}} = 0.968 \pm 0.006\, (68\%\,\mathrm{CL})$ and 
$r < 0.11\, (95\%\,\mathrm{CL})$, and for $\mathcal N\sim 55-65$ the tensor-to-scalar ratio is slightly larger respect to the observations.

For the more general potentials (\ref{ex1bis})--(\ref{ex1tris}) one finds
\begin{equation}
1-n_s=\frac{1+\lambda}{\mathcal N}\,,\quad r=\frac{8\lambda}{\mathcal N}\,.
\end{equation}
The potential $V(\phi)\sim (\phi_0-\phi)$, namely $\zeta=1/2$, in the ``chaotic inflation'' case ($\tilde\alpha=0$) is in agreement with observations for $\mathcal N\simeq 55-60$. The novelity is that this scenario can be also realized by  the
potential $V(\phi)\sim (\phi_0-\phi)^2$ if we take into account the higher contributions associated with $\tilde\alpha$ in our Lagrangian.

For the model in (\ref{ex3}) we get
\begin{equation}
1-n_s=\frac{2}{\mathcal N}
\,,\quad r=\frac{16}{\mathcal N^2}\,,
\end{equation}
similary to Starobinsky-like inflation where $r=12/\mathcal N^2$~\cite{Staro}: these indexes are in good agreement with the Planck satellite data. The result is not surprising, since in scalar field representation the Starobnsky model leads to an exponential potential similar to (\ref{ex3}).
Finally, for the model (\ref{ex5}) we derive
\begin{equation}
1-n_s=\frac{1+\zeta}{\mathcal N}\,,\quad r=\frac{8\zeta}{\mathcal N^{\zeta+1}}\,,
\end{equation}
and a general choice $1<\zeta<2$ leads to viable inflation with $60<\mathcal N<80$. 

We may observe that, given the form of the Hubble parameter, the perturbations during inflation do not depend on $\tilde\alpha$ and $\tilde\beta$, which determine the behaviour of the potential and therefore of the field.

\section{Conclusions}

In this paper, we analyzed several cosmological solutions of a special class of Horndeski models where, despite to the involved form of the Lagrangian, the field equations remain at the second order like in General Relativity. We investigated solutions for inflation, cosmological bounce and finite future-time singularities: all this scenarios may be realized from the model, since it is always possible to reconstruct the suitable forms of the potential which lead to them, showing the versatility of the theory. In the last section, we also revisited the perturbations in FRW space-time and we studied the spectral index and the tensor-to-scalar ratio of our solutions for inflation. Respect to the standard chaotic inflation scenario, it looks that the introduction of higher contributions in the Lagrangian makes viable a larger class of the potentials, like in the case of a quadratic dependence respect to the field.

Our results can be compared with the ones obtained in Ref.~\cite{RGinfl}, where the authors work in the framework
of $f(R,G)$-gravity, $G$ being the Gauss Bonnet four dimensional topological invariant. Without invoking derivatives of the curvature invariants and by using a simple Lagrangian based on the power-law of the Ricci scalar and the Gauss-Bonnet, it is shown how the theory 
can exhaust the whole
curvature budget related to the curvature invariants, and an interesting example of double inflation is presented.

For other works of modified gravity for inflation, see also Refs.~\cite{uno, due, tre, quattro, cinque, sei}.


\end{document}